# Behavioral-clinical phenotyping with type 2 diabetes self-monitoring data

Matthew E. Levine, BA[1]

David J. Albers, PhD[1]

Marissa Burgermaster, PhD[1]

Patricia G. Davidson, DCN, MS, RD, CDE[2]

Arlene M. Smaldone, PhD, CPNP-PC, CDE[3]

Lena Mamykina, PhD[1]

[1]Biomedical Informatics, Columbia University, New York, NY, USA

[2]West Chester University, West Chester, PA, USA

[3]School of Nursing, Columbia University, New York, NY, USA

Corresponding author:

Matthew E. Levine, BA

Research Associate

Department of Biomedical Informatics

Columbia University Medical Center

622 West 168th Street, PH-20

New York, NY 10027; (212) 305-5712

mel2193@cumc.columbia.edu

# ABSTRACT

*Objective*: To evaluate unsupervised clustering methods for identifying individual-level behavioral-clinical phenotypes that relate personal biomarkers and behavioral traits in type 2 diabetes (T2DM) self-monitoring data.

*Materials and Methods*: We used hierarchical clustering (HC) to identify groups of meals with similar nutrition and glycemic impact for 6 individuals with T2DM who collected self-monitoring data. We evaluated clusters on: 1) correspondence to gold standards generated by certified diabetes educators (CDEs) for 3 participants; 2) face validity, rated by CDEs, and 3) impact on CDEs' ability to identify patterns for another 3 participants.

*Results*: Gold standard (GS) included 9 patterns across 3 participants. Of these, all 9 were re-discovered using HC: 4 GS patterns were consistent with patterns identified by HC (over 50% of meals in a cluster followed the pattern); another 5 were included as sub-groups in broader clusers. 50% (9/18) of clusters were rated over 3 on 5-point Likert scale for validity, significance, and being actionable. After reviewing clusters, CDEs identified patterns that were more consistent with data (70% reduction in contradictions between patterns and participants' records).

*Discussion*: Hierarchical clustering of blood glucose and macronutrient consumption appears suitable for discovering behavioral-clinical phenotypes in T2DM. Most clusters corresponded to gold standard and were rated positively by CDEs for face validity. Cluster visualizations helped CDEs identify more robust patterns in nutrition and glycemic impact, creating new possibilities for visual analytic solutions.

*Conclusion:* Machine learning methods can use diabetes self-monitoring data to create personalized behavioral-clinical phenotypes, which may prove useful for delivering personalized medicine.

# BACKGROUND AND SIGNIFICANCE

Chronic disease rates are growing worldwide, making improvements in chronic disease self-management a priority (1). Type 2 diabetes is one of the most prevalent and costly chronic diseases (2); self-management can reduce diabetes-related complications (3,4). However, recent studies demonstrated that individuals have unique glycemic responses to nutrition that depend on complex physiological and contextual factors (5,6). Indeed, it is now acknowledged that successful diabetes treatment and self-management should be tailored to individuals' behaviors as well as their physiology (3,4). Self-monitoring data can be leveraged to personalize self-management strategies; yet, further research is needed to achieve robust personalization (7).

In recent years, *phenotyping* has become a popular approach to personalizing medical treatment to individuals' genetic profile and clinical history. The concept of a phenotype was first introduced to contrast biological traits with their heritability (the genotype) (8,9). More recently, the term has expanded to represent collections of macroscopic observables, not necessarily anchored to a genotype, whose groupings illuminate patterns relevant for understanding human health (10–14).

Phenotyping with clinical data from electronic health records (EHRs) is an area of active investigation (12,14–18). In this context, phenotyping is used to identify clinical patterns that motivate intervention (12,14,15), as well as identify diseases and improve clinical trial recruitment (16,17); it also holds promise for personalizing medical treatments (19,20). Machine learning techniques, such as hierarchical clustering, logistic regression, and neural networks, in conjunction with knowledge engineering approaches, have been used to computationally identify clinical phenotypes (12,15,18,19,21–23).

However, phenotyping approaches can be applied to any macroscopic observable, including behavioral data, which can have special relevance to chronic disease self-management. For example, behavioral phenotypes have been used to describe manifestations of psychiatric disorders (e.g. "a behavioral phenotype is the characteristic cognitive, personality, behavioral, and psychiatric pattern that typifies a disorder" (24)) and to describe eating behaviors, including emotional eating (25) and fat consumption (26–28).

We investigated the feasibility of applying phenotyping techniques to diabetes-related self-monitoring data as a mechanism for personalizing self-management. We examined whether hierarchical clustering, which has proven useful in clinical phenotype construction (29), can identify meaningful relationships between modifiable characteristics of self-management behaviors, for example, inclusion of different macronutrients in an individual's meals, and diabetes-specific biomarkers, like postprandial glycemic response (glucose dynamics that occur shortly after nutritional intake). However, our methods have three important distinctions from previous phenotyping efforts.

*First*, we adopted a particular definition for the term *phenotype*. Previous literature suggests two different ways to conceptualize this phenomenon (11); 1) as the collection of macroscopic traits that define an individual's current state; or 2) as a class of macroscopic trait collections; a disease is a phenotype because it embodies the collections of traits characteristic of individuals with a certain pathophysiological process. Both usages delineate individuals based on macroscopic characteristics, and both can be compatibly mathematized as probability densities over characteristics. However, the first usage is defined relative to an *individual*, whereas the second is defined relative to a *population*. We posit that during exploratory stages of phenotyping, it is essential to first identify and evaluate individual-level phenotypes (specific, scaled feature combinations), which can later form building blocks for general, population-level phenotypes.

Consequently, we focus on characterizing traits within an individual, rather than focusing on population-wide phenomena.

*Second*, unlike previous phenotyping studies that focused on either clinical, e.g. glucose measurements, or behavioral, e.g., consumption of macronutrients in meals, data, we examine integration of these distinct data types into hybrid *behavioral-clinical phenotypes*. This approach has the potential to: 1) help uncover complex and possibly reciprocal relationships *between* clinical and behavioral phenomena, 2) increase phenotype fidelity (30), and 3) form an extensible framework for personalizing behavioral interventions focused on clinical goals.

*Third*, we focus on discovering unnamed, poorly understood phenomena, rather than improving automatic discovery of known collections of traits, and, thus require different validation methods. Standard phenotyping studies are typically validated by some combination of the following steps: *1)* performing checks on face validity, or apparent reasonableness, of identified phenotypes, *2)* assessing agreement with ground truth or "gold standard" (common in clinical domains), *3)* correlating phenotypes to external variables with expected relationships to the phenotypes (common in behavioral domains), and *4)* evaluating the internal phenotype consistency according to relevant criteria, like ratios between distances within and across phenotypes.

Steps 3-4 are not appropriate here because they require specific hypotheses about relationships between variables, absent when phenotyping for personalized discovery. Step 2, gold standard comparison, presents new challenges in the context of self-monitoring data. In clinical phenotyping, gold standards typically focus on presence of a certain condition (binary), are easily expressed quantitatively, and are easily compared with computational phenotypes. In contrast, gold standards for self-management behaviors may involve complex observations and may be expressed qualitatively (e.g. "consuming less than 30% of calories from protein is associated with high glycemic impact"), making them difficult to mathematize for comparison to computationally-generated phenotypes.

To bridge the gap between qualitative gold standards and quantitative analyses, we translate language-based gold standards into mathematical objects (Boolean expressions) and evaluate computed phenotypes against mathematized gold standards. We hypothesize that gold standard inadequacies can be addressed by evaluating mathematized versions for agreement with raw data, and contextualizing computational-phenotype evaluations with gold standard quality. We further hypothesize that expert face validity assessments can additionally compensate for gold standard incompleteness.

This work aims to expand phenotype analyses to the chronic disease management context and outline a methodology for discovering and evaluating unique, personal patterns that elucidate relationships between dietary patterns and glycemic responses in people with type 2 diabetes. This work was guided by four research questions: 1) How can automated phenotyping techniques be used to discover systematic, clinically meaningful trends in self-monitoring data? 2) What feature sets should be selected for phenotype construction from diabetes self-monitoring data? 3) How can behavioral-clinical phenotypes be evaluated and validated empirically? 4) Can behavioral-clinical phenotypes be used to help clinicians better understand patterns in self-monitoring data? By answering these questions, we both define the methodology and work toward improving chronic disease management.

## METHODS

### 1. Data collection and pre-processing

1.1 Participants and data collection

Data were collected from 6 adults with type 2 diabetes recruited from a popular diabetes online community using advertisement on the community's website. During the study, participants recorded photo and text-based descriptions of each meal and its ingredients, along with BG measurements before and 1-3 hours after the meal using a custom smartphone app. The study was approved by the Columbia University Medical Center Institutional Review Board (IRB); all participants digitally signed an informed consent before starting the study.

1.2 Nutrition estimation

After the data collection study, each meal was randomly assigned to one of five registered dietitians (RDs) for evaluation of its macronutrient content. The RDs used each meal's image and textual description to estimate ingredient quantities, and used the USDA National Nutrient Database (https://ndb.nal.usda.gov/ndb/search/list) to determine macronutrient (carbohydrate, fiber, protein, and fat) and calorie composition of each meal.

### 2. Phenotype Identification and Definition

2.1 Defining the Feature Set for Automatic Phenotyping Methods

We worked with two Certified Diabetes Educators (CDEs), with over 40 years of combined clinical and nutritional experience, to identify appropriate features for representing behavioral-clinical phenotypes in diabetes self-monitoring data, including macronutrients in each meal and glycemic impact of each meal. This led to four candidate features sets: 1) percent calories of macronutrients (no BG), 2) grams of macronutrients (no BG), 3) BG impact with percent calories of macronutrients, and 4) BG impact with grams of macronutrients.

2.2 Automated Phenotype Estimation

We used hierarchical clustering to group each participant's meals by their similarity along the selected features (1-4 feature sets above). Hierarchical clustering was selected for its flexibility with respect to distance metric, its ability to detect non-linear and aspherical patterns, its production of hierarchical similarities that remove the need for a prior fixed cluster numbers, and its previous use history in phenotyping studies (31,32). We performed hierarchical clustering using the "hclust" function in R with min-max scaling (to achieve balanced feature scaling with few assumptions on the data distribution), Euclidean distance metric (since we had sufficiently low dimensional continuous variables), and mean linkage criterion (to view each cluster as representing a mean or prototypical meal). Hierarchical clusters were split according to the number of clusters that maximized the Calinski-Harabasz (CH) criterion, a ratio between variance across and within clusters (33). Clusters with fewer than 5 meals were discarded, due to low statistical power and CDEs' recommendations.

**3. Phenotype Evaluation**

We evaluated phenotypes by: 1) comparing with expert-generated gold standards, and 2) assessing face validity. In addition, we examined the impact of phenotypes on experts' ability to assess and understand self-monitoring data.

3.1 Gold Standard Comparison

In order to leverage qualitative gold standards for evaluating computationally-generated behavioral-clinical phenotypes, we performed the following steps: a) generated qualitative gold standards, b) translated the gold standards into computable formats that can be quantitatively evaluated and compared to, c) quantified gold standard quality and correspondence to data, d) selected cluster variables, and e) quantitatively evaluated the degree to which automatic phenotyping rediscovered gold standards.

*3.1.a Qualitative Gold Standard Generation*

To develop gold standards for evaluation, each CDE first worked independently to identify patterns in the self-monitoring datasets, which were formatted into spreadsheets of their specification (features of each individual meal with summary statistics), then CDEs worked together to develop a consensus. Finally, they worked with the research team to ensure their observations used precise language suitable for translation into compound inequality expressions.

For the first three datasets (P1-P3), gold standards were developed twice. The first set was based only on raw data; these gold standards were used for quantitative phenotype evaluation. The second set was developed after CDEs examined both raw data and visualizations from automatic phenotype discovery (parallel coordinate plots of cluster means, (Figure 3)); these gold standards were used to assess how phenotype discovery can influence CDEs' understanding of data. For the remaining datasets (P4-P6), gold standards were created only once, with access to both raw data and phenotype visualizations; these gold standards were also used to assess how phenotype discovery can influence CDEs' understanding of data.

*3.1.b Translating the Written Gold Standard to Computable Inequalities*

We converted expert-generated patterns into compound inequalities. For example, the observation, "Lunch with greater than 45% of calories came from fat, the meals often had high excursion," encodes three components: 1) lunch, 2) percent of calories from fat, and 3) postprandial glycemic increase greater than 50 mg/dl (according to CDEs' definition of "high excursion") and translates to: (meal type == "Lunch") & (calories from fat > 45%) & (blood glucose change > 50 mg/dl).

We then compared translated patterns to the raw data. In several cases when CDEs used ranges of macronutrients to specify patterns, the corresponding compound inequalities showed low correspondence with the data, often because of too narrow specification of ranges. In these cases, we conferred with the CDEs to discuss adjustments to pattern definitions. For example, one observation mapped to no meals when defined on a range of 30-40g of carbohydrates, but mapped to many meals when the carbohydrate range was widened to 25-45g.

*3.1.c Evaluation of the Expert Estimated Gold Standards*

Once we finalized the set of compound inequalities that represented CDEs' consensus of clinical patterns in each dataset, we computed three simple criteria for evaluating gold standard quality: 1) the fraction of all meals that *support* each inequality (met all conditions of the inequality, i.e. true positive rate), 2) the fraction of meals that *contradict* each inequality (met all conditions pertaining to nutrition, but violated conditions pertaining to glucose, i.e. false positive rate), and 3) the total number of inequalities that *over-fit* the data and only apply to 1-2 meals.

*3.1.d Evaluation of Feature Selection*

To identify preferred feature sets for phenotype discovery, we compared gold standards to the clustering output from each of four candidate feature sets. To perform this comparison, we translated the gold standard compound inequalities into a partition of meals. We then quantitatively assessed similarity between discovered partitions and translated gold standard partitions using the Adjusted Rand index (ARI), which quantifies the similarity of partitions, adjusted for chance element placement (34,35). We used an online tool (36) to compute ARI and estimate its 95% confidence interval (37). We selected the feature set that most often maximized ARI for downstream quantitative analysis of phenotype discovery.

*3.1.e Evaluation of Automatically Estimated Phenotypes Against the Gold Standard*

We compared phenotypes to the gold standard for P1-3 generated by CDEs using only raw self-monitoring data using two different metrics.

First, we identified clusters that likely contained enough information to rediscover each gold standard observation, specifically, observations whose meals comprise at least 50% of a single cluster. We used the number of *rediscovered whole-phenotype observations*, and the fraction of meals accounted for in this way, to evaluate the degree to which hierarchical clustering can be used to automatically rediscover clinical patterns in self-monitoring data.

Second, we identified clusters that likely represented a special case of a more general pattern identified by CDEs. These *rediscovered partial phenotype observations* were defined as cases when >50% of meals from the observation belong to the cluster. This provided an indication of how similarly the algorithm split the data compared to expert observations.

3.2 Face Validity

In order to evaluate face validity of the identified phenotypes, CDEs were shown visualizations of clustering results (parallel coordinate plots of cluster means, (Figure 3)), along with raw self-monitoring data. They were shown two versions of clustering results from two feature sets (blood glucose change with grams of macronutrients and blood glucose with percent calories from macronutrients). CDEs first identified the preferable feature set for the given dataset based on two sets of output, then rated each cluster in the preferred output from 1-5 (1: not-at-all, 5: absolutely) in terms of whether the cluster was: 1) valid based on the data, 2) clinically significant, and 3) actionable; they also provided a rationale for their responses.

In order to specifically evaluate our ability to uncover hard-to-detect phenotypes, we separately reported face validity metrics for clusters that did not align with any gold standard observations.

3.3 Evaluation of the Impact of Automatically Estimated Phenotypes on Expert Analysis

In order to evaluate whether results from automatic phenotyping can support CDEs' ability to understand and analyze diabetes self-monitoring data, CDEs generated new gold standards using both the raw self-monitoring data and visualizations of clustering results (parallel coordinate plots of cluster means, (Figure 3)). For P4-6 datasets, this approach was used upon the CDEs' first data viewing; for P1-3, this was done after an original gold standard was created using only raw data. We used our three gold standard quality metrics (fractions of meals that *support*, *contradict*, and

*over-fit* the gold standard observations) to compare the quality of gold standards generated with and without phenotype visualizations.

**RESULTS**

Self-monitoring data collected from six individuals is summarized in Table 1. Figure 1 illustrates the unique behavioral-clinical patterns in individuals' data.

**4. Automatic Phenotype Identification**

Figure 2a-b depicts the results from hierarchical clustering of nutritional content of meals from the P2 and P3 dataset using the grams of macronutrients feature set. Some trends are readily visible (e.g. for P2, low carbs and low fiber align with smallest glucose increases). Figure 2c-d show the Calinski-Harabasz optima, which was used to select an appropriate number of clusters; for P2 (fig. 2b), 10 clusters were selected, and clusters with fewer than 5 meals were discarded, leaving 4 analyzed clusters. The resulting clusters for P2 and P3 are summarized in figure 3 by showing parallel coordinate plots of each cluster mean, and all cluster summaries are provided in supplementary figure 2.

**5. Phenotype Evaluation**

5.1 Gold Standard Comparison

*5.1.a-b Gold standard creation and translation*

CDEs generated 23 total observations across three participants (12 for P1, 7 for P2, and 4 for P3, see Tables 1-3 in supplementary materials). Fourteen observations were excluded from analysis, because they described nutrition or glucose, but not both jointly (7 in P1, 1 in P2, 2 in P3) or because they specified unanalyzed variables, including time between meals and specific ingredients (3 in P1, 1 in P2), leaving nine gold standard observations for analysis in P1-3. Of these, six specified meal-type and six specified more than 1 macronutrient. Carbohydrate was the most frequently specified macronutrient (5/9) and fiber the least (2/9). Observations were translated into Boolean queries (supplementary tables 1-3).

*5.1.c Evaluation of the Expert Estimated Gold Standards*

Of the nine analyzed observations, two were over-fit (only fit 1 meal; 1 in P2, 1 in P3), and two were contradicted by the data more often than they were supported (1 in P2, 1 in P3). For example, P3 observation three states "higher inclusion of fat (over 40%) leads to high excursion" and fits with 30% of P3's meals. However, 50% of P3's meals are meals with more than 40% fat and low glycemic excursions under 50 mg/dl. Observation redundancy was also observed (e.g. observation 6 in P2 subsumes 85% of observation 2). The remaining observations were generally consistent with the data (supported more often than contradicted) and presented good coverage (over 10% of the meals in the corresponding datasets). Five observations (22%) initially fit zero meals, but iterative adjustments substantially improved their fit with the data. Table 2 summarizes gold standard observation quality; supplementary tables 1-3 contain the raw observations, their Boolean translations, and their evaluation statistics.

*5.1.d Evaluation of Feature Selection*

We evaluated similarity between gold standard partitions and clusters of P1-3 using the adjusted Rand Index (ARI) (Table 3). Clustering with "BG with grams of macronutrients" feature set maximized ARI for two of three participants (P1 and P3), and was second-best for P2.

*5.1.e Evaluation of Automatically Estimated Phenotypes Against the Gold Standard*

Comparisons of automatically generated clusters (using ARI-selected feature set) with CDE-generated observations ("observations" refer to the gold standards) showed that five of nine total observations were rediscovered as whole-phenotypes through clustering (comprised at least 50% of meals in a cluster) (1 in P1, 4 in P2). The four remaining observations that comprised minority portions of clusters exhibited poor correspondence with the data—they either represented less than 10% of the data or were frequently contradicted. However, all low-frequency CDE-observations were rediscovered as partial phenotypes, or sub-clusters of our phenotypes. For example, 100% of observation of eight meals in P3 were placed into cluster two. Two of the remaining observations were trivially sub-clustered (each corresponded to a one-meal sub-cluster). All results are reported in Supplementary Tables 4-6.

5.2 Face Validity

*5.2.1 Evaluation of Expert-based Face Validity of Automatic Phenotyping*

Table 4 reports and summarizes CDEs' phenotype face validity survey responses. Overall, the CDEs rated 50% (9 of 18) of identified clusters over three (on a 1-5 Likert scale) for validity, significance, and being actionable. Four of the nine remaining lower-rated clusters lacked meaning and contained mostly noise—they scored below three in all three categories, and did not align with any CDE observations. Three clusters scored very highly on validity, but lacked clinical significance; one was found not actionable, and one lacked clinical significance. The primary reason for low-scoring clusters was high within-cluster variance of blood glucose or macronutrients. Clusters that grouped meals with moderate glycemic impact (20-50 mg/dl change) were typically considered not actionable and lacking clinical significance.

*5.2.2 Evaluation of Automatically Estimated Novel Phenotypes:*

While all nine analyzed gold standard observations from P1-3 datasets were rediscovered as either partial or whole phenotypes, we also assessed the five novel clusters that did not align with any gold standard observation and may contain valuable information missing from the gold standard.

Two of five novel clusters were rated three or above on a five-point Likert scale in all three categories of validity, indicating their reasonableness as novel discoveries. Three novel clusters (one in P2, two in P6) received low face validity scores, and were described as having low validity based on the data, not clinically significant, and not actionable.

5.3 Evaluation of the Impact of Automatically Estimated Phenotypes on Expert Analysis

Table 2 shows that observations generated with cluster visualizations (P4-6, reported in supplementary tables 7-9) were supported by fewer total meals than those generated with only raw data (P1-3), but were contradicted by the data less frequently. Two of the seven analyzed observations were over-fit (2 in P4), and only 1 observation was contradicted more often than it was supported. Table 2 also shows that observations generated with cluster visualizations were supported by fewer meals in P1 and P2 but reduced the contradiction rate by over 70% for P2 and P3 (details in supplementary tables 7-9).

**DISCUSSION**

In this study, we applied phenotyping methods to the context of personal health data and diabetes self-management. Our research questions focused on applicability of the phenotyping techniques to self-monitoring data, appropriate feature sets, appropriate methods for validating such phenotypes, and their impact on CDEs' ability to better understand patterns in patients' data.

Overall, we found that 50% (9 out of 18) of clusters discovered using hierarchical clustering had high correspondence with the gold standard; moreover, this approach reproduced five of nine CDE-generated observations, and identified generalizations of the remaining four observations. Given the large volume of possible partitions of the datasets used in this study, this indicates a substantial correspondence between the phenotypes and the gold standards. These results suggest that clustering approaches can serve as building blocks for establishing individual- and, possibly, population-level behavioral-clinical phenotypes, and may form important analytical structures in behavioral informatics interventions. Moreover, more than half of all clusters received high ratings for being clinically valid, significant, and actionable. This suggests that such clusters have the potential to provide useful assistance in identifying significant traits in patients' self-management behaviors, which could be used to personalize their self-management strategies.

In regards to the feature set selection, we found that including blood glucose with grams of macronutrients most often optimized the ARI partition similarity metric. This was partially consistent with perceptions of CDEs, who suggested including glycemic impact as a clustering variable. However, these findings contradicted their expectation that proportion of calories contributed by macronutrients better explains nutritional-glycemic trends than absolute macronutrient weights. This finding suggests that including health biomarkers as part of clustering is a fruitful alternative to clustering on either behaviors (nutrition) or clinical outcomes, as was done in previous efforts to phenotype behaviors (13). It also indicates a potential need for healthcare professionals to select different clustering features on a case-by-case basis.

Moreover, the study suggested that exposing CDEs to computationally generated clusters led them to generate observations with higher specificity and lower sensitivity (i.e. the CDEs gave a more conservative account of patterns present in the data). When viewing phenotype visualizations, CDEs identified fewer patterns overall, and accounted for fewer meals in their observations (reduced sensitivity); however, each pattern was contradicted less often and typically had more support from the data (higher specificity).

At the same time, this study identified open questions in regards to applying computational approaches to discovering behavioral-clinical phenotypes. For example, evaluation of such phenotypes through a gold-standard validation pipeline presented considerable difficulties and necessitated modifying traditional evaluation approaches. The availability of gold standards was vital to our evaluation, as they imposed a domain-specific quality metric and allowed for scalable model selection. Because gold standards for behavioral-clinical phenotypes predominantly exist as qualitative expert knowledge, innovations were necessary to bridge the gap between the qualitative and quantitative descriptions of identified patterns. Additional methods were necessary for assessing gold standard quality and contextualizing their comparisons. Our findings suggest that further innovations, beyond manual, iterative translation of qualitative statements into quantitative expressions, are necessary for efficient, reliable use of gold standards for validating complex, unnamed phenomena. While validation with an imperfect gold standard poses fundamental limitations, prospective validation experiments, including n=1-trials and self-experimentation framework (38,39), and other task-based validity approaches may be better suited to validating behavioral-clinical phenotypes.

In addition, while this study suggested that computational methods used in traditional phenotyping research can be applied to non-clinical data in order to identify behavioral-clinical phenotypes, questions remain as to how these phenotypes can be used for personalizing self-management strategies. Moreover, further research is needed to identify appropriate ways to design new informatics interventions that use personalized self-management strategies to help individuals with chronic conditions improve their health.

**LIMITATIONS**

Our cohort included 6 people; these results may not be generalized to all people with type 2 diabetes. Moreover, most participants were well controlled with self-reported HbA1c values around 6% and exhibited low variance in nutritional intake; the method may generalize differently in patients with less controlled diabetes, or with greater variability in nutritional intake. Definitions for rediscovered whole and sub-phenotype patterns were chosen as simple heuristics, and are not complete assessments of correspondence between gold standard and automatic clusters. In general, comparison becomes difficult when the gold standard is not a true partition (contains multiple mappings and null mappings). Nevertheless, these metrics provide first-order approximations of similarity.

## CONCLUSION

We have developed an implementation of hierarchical clustering and an evaluation methodology for identifying and validating clinically meaningful behavioral-clinical structures and the features used to define them from diabetes self-monitoring data. The evaluated phenotyping approach and its population-level successors create exciting opportunities to support individuals in exploring and improving their health. Our long-term vision is to use self-monitoring data to uncover important behavioral-clinical phenotypes that can shed light on treatment practices and help identify specific, personalized self-management strategies. In this way, behavioral-clinical phenotyping can form the analytical backbone for informatics interventions that take a holistic, data-driven approach to personalized medicine.

## ACKNOWLEDGMENTS

All authors made substantial contributions to the conception and design of the work. ML wrote the original draft and performed all clustering and statistical analyses. ML, LM, MB, and DJA designed the evaluation study. PGD and AMS created gold standards and provided expert evaluations. All authors revised the draft for important intellectual content.

## COMPETING INTERESTS

The authors have no competing interests to declare.

## FUNDING

This work was funded in part by the Robert Wood Johnson Foundation grant 73070, National Institutes of Health (NIH) National Institute of Diabetes and Digestive and Kidney Diseases (NIDDK) grant R56DK113189, NIH National Library of Medicine (NLM) grant R01 LM06910, NIH NLM grant R01 LM009886, and NIH National Human Genome Research Institute (NHGRI) grant U01 HG008680.

**FIGURES**

a)

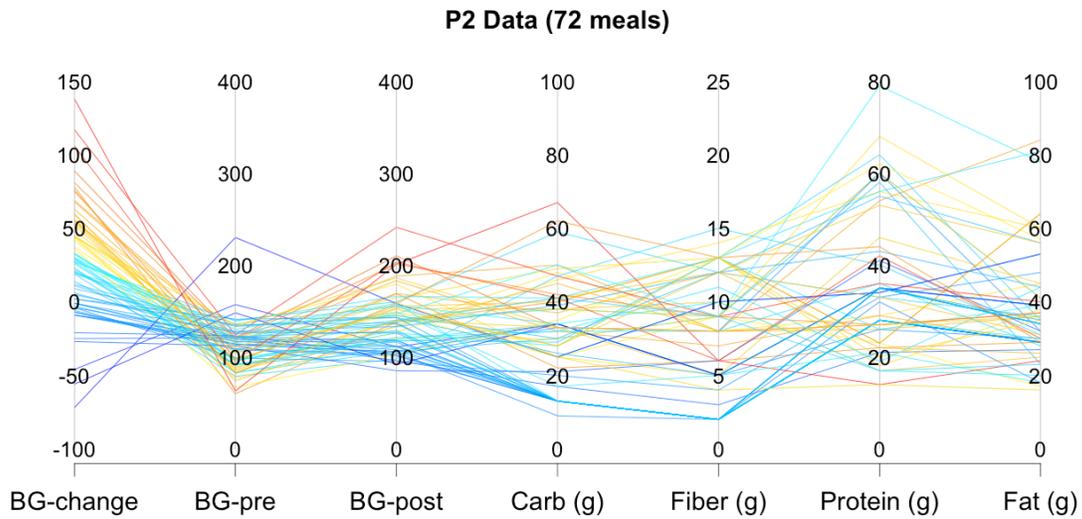

b)

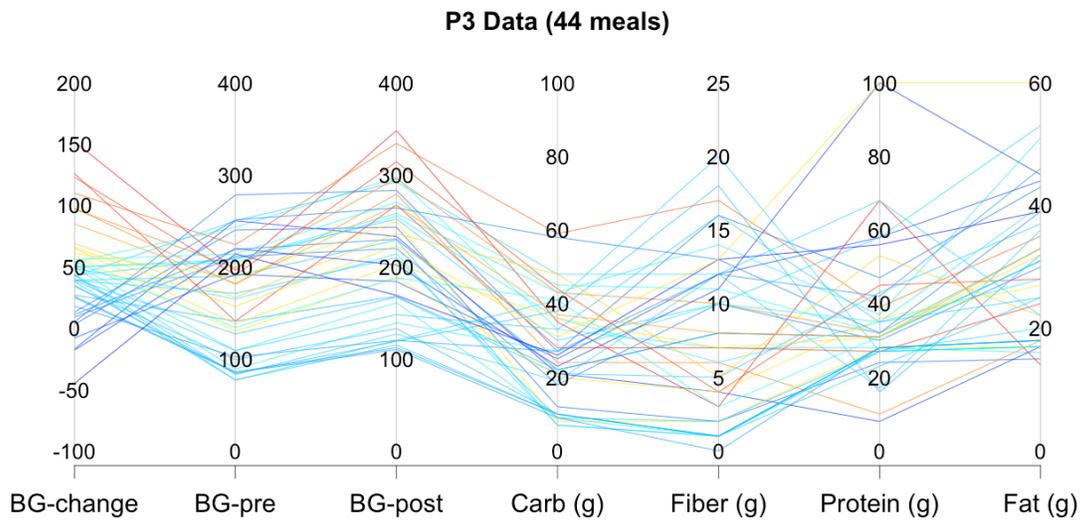

**Figure 1:** Parallel coordinate plots of behavioral-clinical self-monitoring data from P2-P3. Each line represents a single meal, and each vertical axis represents a behavioral or clinical variable (labeled at the bottom of each plot). The line color indicates the degree of glucose change, where red is highest and blue is lowest. These plots demonstrate the heterogeneity of behavioral-clinical patterns within and across participants.

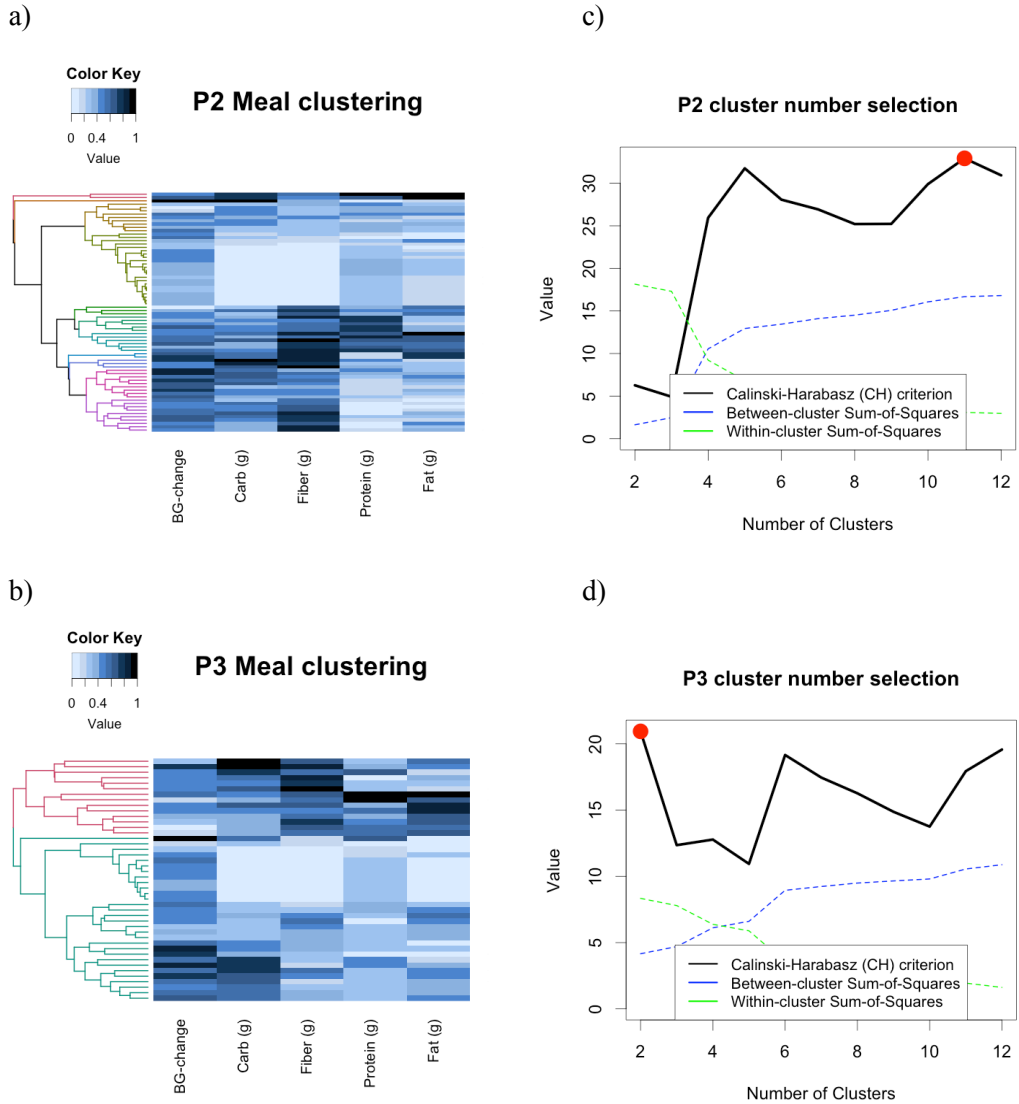

**Figure 2:** Raw clustering analysis for P2-P3. In figures (a) and (b), we show heatmaps of normalized behavioral-clinical variables for P2 and P3, where each row represents a meal. The data are max-min normalized by each column, so that the darkest row in a column represents the meal with the largest value in that column. The dendrogram tree on the left axis indicates the hierarchical similarities between groups of meals. The blocked sections of the dendrogram indicate how the tree is cut when selecting a particular number of clusters (for P2, 11; for P3, 2). In figures (c) and (d), we plot the Calinski-Harabasz (CH) criterion for 1-12 clusters, and find it takes a maximum at 11 clusters for P2 and at 2 for P3, indicating an advantageous balance between internal consistency of clusters and their dissimilarity from each other. The CH criterion is related to the ratio of average between- and within-cluster sums of squared distances, both of which we plot for comparison. Note that both within-cluster and between-cluster variance saturates as cluster number increases. As expected, within-cluster variance decreases and between-cluster variance increases as cluster number increases.

a)

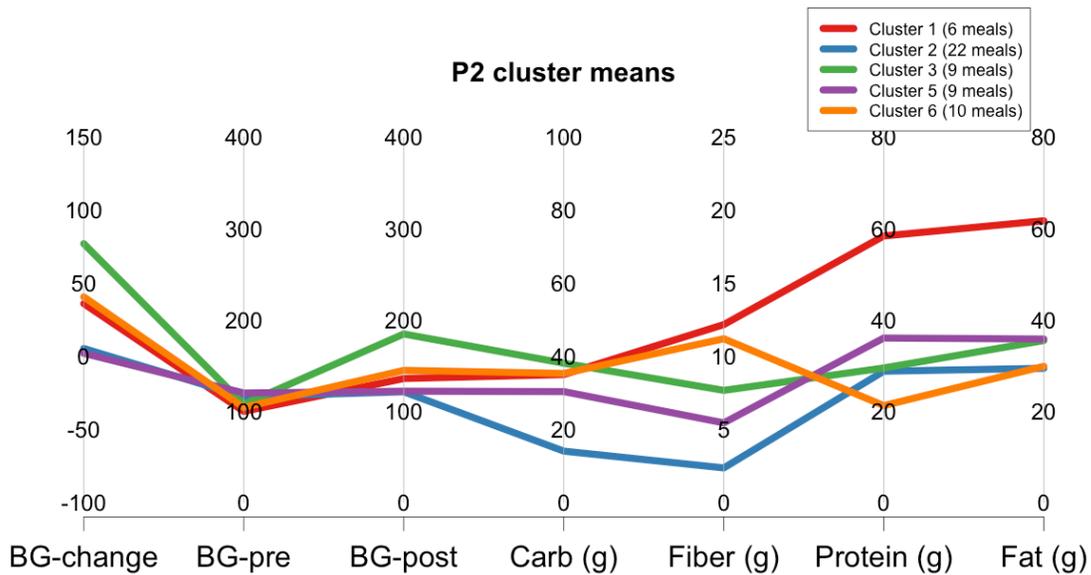

b)

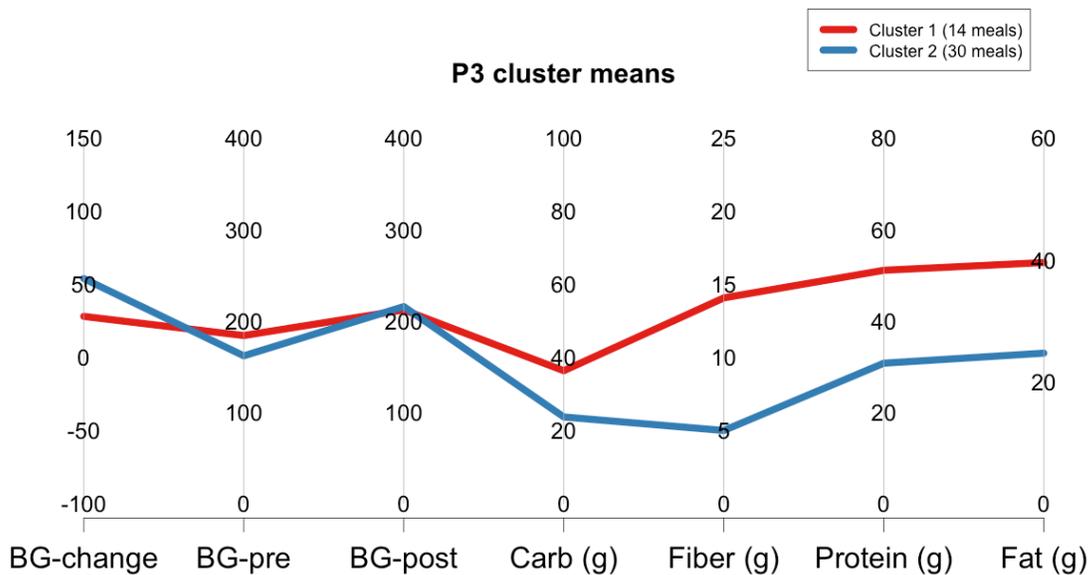

**Figure 3:** Parallel coordinate plots of cluster means from P2 (a) and P3 (b). Each line represents the mean values for a single cluster. Clustering was performed using the blood glucose with grams of macronutrients feature set, and clusters with fewer than 5 meals were excluded. We observe that each participant exhibits different personal behavioral-clinical phenotypes. P2 clearly has a group of meals with moderate-high glycemic impact that are high in protein and fat (red). P3 shows a phenotype where lower carbohydrates, in conjunction with lower fiber, protein, and fat lead to higher glycemic impacts. Note that CDEs were shown similar visualizations when generating gold standards, but they received more detailed plots with additional axes to indicate other variables they were interested in.

# TABLES

| Participant ID | **P1** | **P2** | **P3** | **P4** | **P5** | **P6** |
|---|---|---|---|---|---|---|
| Age | 40−50 | 40−50 | 40-50 | 50-60 | 50-60 | 50-60 |
| Disease Status | T2D | T2D | T2D | T2D | T2D | T2D |
| Diabetes Medications | Oral agents | Oral agents | None | Oral | Oral | Insulin, |
| # of glucose measurements | 304 | 211 | 89 | 118 | 86 | 70 |
| # of meals recorded | 120 | 76 | 45 | 60 | 48 | 36 |
| # of meals with associated pre- and post-meal glucose | 101 | 72 | 44 | 56 | 40 | 34 |
| # of days measured | 27 | 28 | 51 | 36 | 32 | 21 |
| Mean measured glucose (mg/dL) | 113±25 | 127±32 | 192 ± 65 | 113±16 | 97±19 | 121 ± 31 |
| Mean measured pre-meal glucose (mg/dL) | 97±15 | 111±25 | 169 ± 57 | 105±9 | 87±11 | 111 ± 21 |
| Mean measured post-meal glucose(mg/dL) | 126±26 | 139±32 | 214 ± 65 | 124±16 | 110±19 | 132 ± 36 |

**Table 1:** Participant Self-Monitoring Data Description

| Particip ant ID | Data available during gold standard generation | Number of analyzed expert observations | Number of over-fit observations | Number of contradictory observations | % meals supported by observations | % meals contradicted by observations |
|---|---|---|---|---|---|---|
| P1 | Raw self-monitoring | 2 | 0 | 0 | 31% | 7% |
| P2 | Raw self-monitoring | 5 | 1 | 1 | 74% | 23% |
| P3 | Raw self-monitoring | 2 | 1 | 1 | 32% | 52% |
| P4 | Raw self-monitoring & cluster-vis. | 4 | 2 | 1 | 27% | 23% |
| P5 | Raw self-monitoring & cluster-vis. | 2 | 0 | 0 | 25% | 13% |
| P6 | Raw self-monitoring & cluster-vis. | 1 | 0 | 0 | 29% | 0% |
| P1 | Raw self-monitoring & cluster-vis. | 4 | 0 | 0 | 24% | 14% |
| P2 | Raw self-monitoring & cluster-vis. | 2 | 0 | 0 | 43% | 4% |
| P3 | Raw self-monitoring & cluster-vis. | 5 | 0 | 0 | 59% | 14% |

**Table 2:** Here, we summarize our evaluation of gold standard quality. We have 1 row for each gold standard, and P1-3 appear twice because they were each evaluated twice separately using different source data (raw self-monitoring data refers to excel spreadsheets made to CDEs' specifications, "cluster-vis" refers to parallel coordinate plots of cluster means (c.f. Figure 3). The number of *over-fit* and *contradictory* gold standard observations provide a measure of over-generalization and inconsistency with the data, respectively. The fraction of meals *supported* and *contradicted* by the gold standard observations indicates the overall rates of true positives and false negatives in the gold standard.

| Participant ID | Grams of Macronutrients | %-Calories from Macronutrients | Blood Glucose & Grams of Macronutrients | Blood Glucose & %-Calories from Macronutrients |
| --- | --- | --- | --- | --- |
| P1 | 0.018 (0.000-0.232) | 0.273 (0.128-0.408) | **0.495** (0.215-0.763) | 0.00 (0.00-0.00) |
| P2 | 0.472 (0.225-0.735) | **0.611** (0.373-0.870) | 0.549 (0.307-0.810) | 0.544 (0.319-0.779) |
| P3 | 0.074 (0.000-0.253) | 0.086 (0.000-0.324) | **0.344** (0.000-1.000) | 0.130 (0.000-0.471) |

**Table 3:** Similarity between gold standard observations and cluster results using different feature sets. Partition similarity based on Adjusted Rand Index (ARI) and its 95% confidence interval. Note that ARI ranges from 0-1, with 0 indicating no partition similarity and 1 indicating perfect partition similarity—higher ARI indicates greater similarity to gold standard. We note that ARI for P1 and P3 is maximized using the blood glucose and grams of macronutrients feature set.

|  | Valid (1-5) | Significant (1-5) | Actionable (1-5) | CDEs' feature preference |
|---|---|---|---|---|
| **P1** | 3.5 | 4 | 4 | BG with Percent-calories from macronutrients |
|  | 3.5 | 4 | 4 |  |
| **P2** | 3.75 | 4 | 4 | BG with grams of macronutrients |
|  | 3.5 | 3.5 | 4 |  |
|  | 5 | 4 | 4 |  |
|  | 2 | 2 | 1.5 |  |
|  | 4 | 3.5 | 2 |  |
| **P3** | 4 | 4 | 4 | BG with Percent-calories from macronutrients |
|  | 4 | 3 | 4 |  |
|  | 2 | 4 | 2 |  |
| **P4** | 1 | 1 | 1 | BG with Percent-calories from macronutrients |
|  | 4 | 2 | 2 |  |
| **P5** | 1.5 | 1.5 | 1 | BG with Percent-calories from macronutrients |
|  | 4 | 2.5 | 4 |  |
|  | 4 | 4 | 3.5 |  |
| **P6** | 1 | 1 | 1 | BG with grams of macronutrients |
|  | 4 | 4 | 4 |  |
|  | 4 | 4 | 4 |  |
|  | 1 | 1 | 1 |  |
| **Total mean** | **3.14** | **3.0** | **2.89** |  |

**Table 4:** Face validity Likert scores (on a scale of 1-5, strongly disagree to strongly agree). For each participant, CDEs were presented with clustering results from two different feature sets (BG with grams of macronutrients, BG with percent-calories from macronutrients), and CDEs selected which feature set's results they preferred and would like to evaluate. They elected to evaluate BG with percent-calories from macronutrients for 4/6 participants even in cases where ARI indicated that the other feature set was more similar to their gold standard.

**SUPPLEMENT**

| Observation # | Text Description | Boolean Query | %-meals in agreement | %-meals with opposite glycemic impact |
|---|---|---|---|---|
| 8 | When breakfasts had higher proportion of carbohydrates (more than 40%) and lower in protein (less than 40%), there is a higher differential (over 100) in 1 hour | Meal-type == "Breakfast" calories-from-carbs > 35% calories-from-protein < 35% BG-change > 50 | 10% | 6% |
| 9 | Lunch: high inclusion of protein (over 55%) with generally mild glycemic impact | Meal-type == "Lunch" calories-from-protein > 55% BG-change < 50 | 1% | 0% |
| 9 revised-1 | | Meal-type == "Lunch" calories-from-protein > 30% BG-change < 50 | 6% | 0% |
| 9 revised-2 | | calories-from-protein > 30% BG-change < 50 | 22% | 1% |
| | TOTAL | 32 | 31% meals accounted for by observations | 7% meals contradict an observation |

**Supplementary Table 1:** P1 Gold Standard evaluation

**Supplementary Table 2:** P2 Gold Standard evaluation

| Observation # | Text Description | Boolean Query | %-meals in agreement | %-meals with opposite glycemic impact |
|---|---|---|---|---|
| 2 | Moderate amount of carbohydrate (around 30g-40g) combined with higher fiber (around 9-10g or higher) lead to more moderate impact (under 50) | 25g < carbs < 45g<br>fiber > 8g<br>BG-change < 50 | 18% | 11% |
| 3 | High carbohydrate meals (over 40g) that have more even inclusion of other macronutrients have more moderate impact (around 50 or below; particularly for dinner) but low certainty because only few cases | carbs > 35g<br>calories-from-carbs > 25%<br>calories-from-protein > 25%<br>calories-from-fat > 25%<br>BG-change < 50 | 1% | 3% |
| 4 original | Breakfast generally minimal excursion (carbs between 13 and 15, which is lower than for other types of meals because of the proportionately higher fiber, around 9g or higher, as compared to other meals) | Meal-type == "Breakfast"<br>10g <carbs< 20g<br>fiber > 8g<br>BG-change < 50 | 0% | 0% |
| 4 revised | Breakfast generally minimal excursion (carbs between 13 and 15, which is lower than for other types of meals because of the proportionately higher fiber, around 9g or higher, as compared to other meals) | Meal-type == "Breakfast"<br>10g < carbs < 20g<br>BG-change < 50 | 25% | 0% |
| 5 | Lunch when greater than 45% of calories came from fat, the meals often had high excursion | Meal-type == "Lunch"<br>calories-from-fat > 45%<br>BG-change > 50 | 17% | 10% |
| 6 | Dinner the lunch trend of high proportions does not hold. At dinner, similarly high proportions of fat and protein did not lead to high excursions, potentially because of higher fiber (10g and above), but low certainty because of variability | Meal-type == "Dinner"<br>(calories-from-fat > 45% OR calories-from-protein > 45%)<br>BG-change < 50 | 29% | 7% |
| | TOTAL | 53 | 74% of meals accounted for by observation | 23% meals contradict an observation |

**Supplementary Table 3:** P3 Gold Standard evaluation

| Observation # | Text Description | Boolean Query | %-meals in agreement | %-meals with opposite glycemic impact |
|---|---|---|---|---|
| 3 | Overall: higher inclusion of fat (over 40%) leads to high excursion | calories-from-fat > 40% <br> BG-change > 50 | 30% | 52% |
| 4 | Dinner: higher carb (30%) and lower fat (under 40%) lead to low excursion | Meal-type == "Dinner" <br> calories-from-carbs > 30% <br> calories-from-fat < 40% <br> BG-change < 50 | 2% | 0% |
|  | TOTAL | 14 | 32% meals accounted for by observations | 52% of meals contradict an observation |

**Supplementary Table 4:** P1 Cluster-Gold-Standard alignment

| clusterID | # of meals | % of a given cluster that satisfies: | | | | |
| --- | --- | --- | --- | --- | --- | --- |
| | | Any Observation | OBS 3 | OBS 4 | OBS 5 | OBS 6 |
| 1 | 6 | 50% | | | 50% | |
| 2 | 25 | 56% | 56% | 40% | | |
| 3 | 22 | 0% | | | | |
| 4 | 42 | 14% | | | | 14% |

Note: "Observation" is abbreviated as "OBS"

**Supplementary Table 5:** P2 Cluster-Gold-Standard alignment

| clusterID | # of meals | % of a given cluster that satisfies: | | | | | |
| --- | --- | --- | --- | --- | --- | --- | --- |
| | | Any Observation | OBS 2 | OBS 3 | OBS 4 | OBS 5 | OBS 6 |
| 1 | 6 | 83% | 67% | | | | 83% |
| 2 | 22 | 91% | | | 81% | | 9% |
| 3 | 9 | 78% | | | | 78% | |
| 5 | 9 | 11% | | | | 11% | |
| 6 | 10 | 90% | 50% | | | 20% | 60% |

Note: "Observation" is abbreviated as "OBS"

| clusterID | # of meals | % of a given cluster that satisfies: | | |
| --- | --- | --- | --- | --- |
| | | Any Observation | OBS 3 | OBS 4 |
| 1 | 14 | 21% | 14% | 7% |
| 2 | 30 | 37% | 37% | |

**Supplementary Table 6:** P3 Cluster-Gold-Standard alignment

Note: "Observation" is abbreviated as "OBS"

| Observation # | Text Description | Boolean Query | %-meals in agreement | %-meals with opposite glycemic impact |
|---|---|---|---|---|
| 2 | The highest glycemic excursions come after breakfast | mealType == "Breakfast" BGchange > 80% − quantile | 13% | 5% |
| 3 | As carbohydrate content goes up (over 22%), and fat goes up (over 40%) fiber content goes down (below 8g), which contributes to higher excursion | calories-from-carbs > 22% calories-from-fat > 40% fiber < 8g BGchange > median−BGchange | 16% | 11% |
| 5 | When fiber is 8g or above, and carbs are below 22% and fat is above 60% there is lower excursion | fiber ≥ 8g calories-from-carbs < 22% calories-from-fat > 60% BGchange < median−BGchange | 2% | 0% |
| 6 | Higher fat content in the morning (breakfast) is keeping BG excursion low despite lower fiber (below 8g) | mealType == "Breakfast" calories-from-fat > 40% fiber < 8g BGchange < median−BGchange | 2% | 7% |
| | TOTAL | 15 | 27% meals accounted for by observations | 23% of meals contradict an observation |

**Supplementary Table 7:** P4 Expert Observations

| Observation # | Text Description | Boolean Query | %-meals in agreement | %-meals with opposite glycemic impact |
|---|---|---|---|---|
| 2 | For dinner meals, meals that have 40-60% fat have mild impact | mealType == "Dinner"<br>40% < calories-from-fat < 60%<br>BGchange < 50 | 18% | 13% |
| 3 | For breakfasts, meals that have over 40% of protein, and about 35% of carbs have mild impact | mealType == "Breakfast"<br>30% < calories-from-carbs < 40%<br>calories-from-protein > 40<br>BGchange < 50 | 8% | 0% |
| | TOTAL | 10 | 25% meals accounted for by observations | 13% of meals contradict an observation |

**Supplementary Table 8:** P5 Expert Observations

| Observation # | Text Description | Boolean Query | %-meals in agreement | %-meals with opposite glycemic impact |
|---|---|---|---|---|
| 4 | Minimum of 5g of fiber and fat content between 40%-60% then BG impact < 50 (based on cluster 4) | fiber >= 5g<br>35% < calories-from-fat < 65%<br>BGchange < 50 | 29% | 0% |
| | TOTAL | 10 | 29% meals accounted for by observations | 0% of meals contradict an observation |

**Supplementary Table 9:** P6 Expert Observations

| Observation # | Text Description | Boolean Query | %-meals in agreement | %-meals with opposite glycemic impact |
|---|---|---|---|---|
| 3 | The majority of high impact meals are breakfasts (have the highest BG excursion and highest post-meal) | Meal-Type == "Breakfast" BG-change > 50 | 15% | 11% |
| 4 original | High carbohydrate (over 40%) in breakfast meals, combined with low protein (20%) lead to high excursions | BG-change > 50 calories-from-carbs > 0.4 calories-from-protein < 0.2 meal-Type == "Breakfast" | 5% | 5% |
| 4 revised | | BG-change > 50 calories-from-carbs > 0.35 calories-from-protein < 0.25 meal-Type == "Breakfast" | 10% | 6% |
| 5 | High fiber (over 15g) and moderate carb (20-40%) leads to low impact (lower than 50) | BG-change < 50 0.2 < calories-from-carbs < 0.4 fiber > 15g | 3% | 0% |
| 6 | In meals with fat around 35-40% and low protein (under 10g), mild impact (this pattern seems to be only holding in some clusters, but not others) | BG-change < 50 0.35 < calories-from-fat < 0.4 protein < 10g | 3% | 1% |
| 6 revised | | BG-change < 50 0.3 < calories-from-fat < 0.45 protein < 10g | 6% | 3% |
| | TOTAL | 24 | 24% meals accounted for by observations | 14% of meals contradict an observation |

**Supplementary Table 10:** P1 Expert Observations (Post-Cluster)

| Observation # | Text Description | Boolean Query | %-meals in agreement | %-meals with opposite glycemic impact |
|---|---|---|---|---|
| 1 | Higher fiber does not seem to have a positive impact on BG (meals with low fiber, lower than 5g, have low glycemic impact) | BG-change < 50<br>fiber < 5g | 29% | 0% |
| 2 original | High fiber (over 10g) and high fat (45-60%) together lead to low impact | BG-change < 50<br>fiber > 10g<br>45% < calories-from-fat < 60% | 10% | 3% |
| 2 revised | | BG-change < 50<br>fiber > 10g<br>40% < calories-from-fat < 65% | 14% | 4% |
| | TOTAL (72 meals) | 31 meals accounted for | 43% meals accounted for by observations | 4% of meals contradict an observation |

**Supplementary Table 11:** P2 Expert Observations (Post-Cluster)

| Observation # | Text Description | Boolean Query | %-meals in agreement | %-meals with opposite glycemic impact |
|---|---|---|---|---|
| 1 | High proportion of fat (on average most meals have over 50%) | calories-from-fat > 50% | 64% | 11% |
| 4 | The highest BG excursions (50 to 120) came from meals with higher total number of grams of carb (from 20 to 45g) and highest proportion of carbs and lower fiber (less than 10g) even with high fat | calories-from-carbs > median carbs > 20g fiber < 10g BG-change > 50 | 20% | 7% |
| 5 | Higher fat (over 40g) and higher fiber (over 10g) lead to lower glycemic excursion | fat > 40g fiber > 10g BG-change < 50 | 9% | 3% |
| 6 | Higher fiber (over 10g) leads to lower excursion (particularly at dinner) | Meal-Type == "Dinner" fiber > 10g BG-change < 50 | 25% | 3% |
| 7 | Breakfast: proportion of carbs is the lowest, and protein is the highest, and BG excursion is moderate | Meal-Type == "Breakfast" calories-from-carbs < 30%−quantile calories-from-protein > 70%−quantile BG-change < 50 | 16% | 7% |
| 8 | Lunch: the highest BG change, grams of carbs is the highest (around 40g), proportion of carbs is the highest (30%), high fat (40%) and lowest protein (20%), and lower fiber (under 8g) | Meal-Type == "Lunch" calories-from-carbs > 30% calories-from-fat > 40% calories-from-protein < 20% fiber < 8g carbs > 40g BG-change > 50 | 0% | 0% |
| | TOTAL | 26 | 59% meals accounted for by observations | 14% of meals contradict an observation |

**Supplementary Table 12:** P3 Expert Observations (Post-Cluster)

**Supplementary Figure 1.**

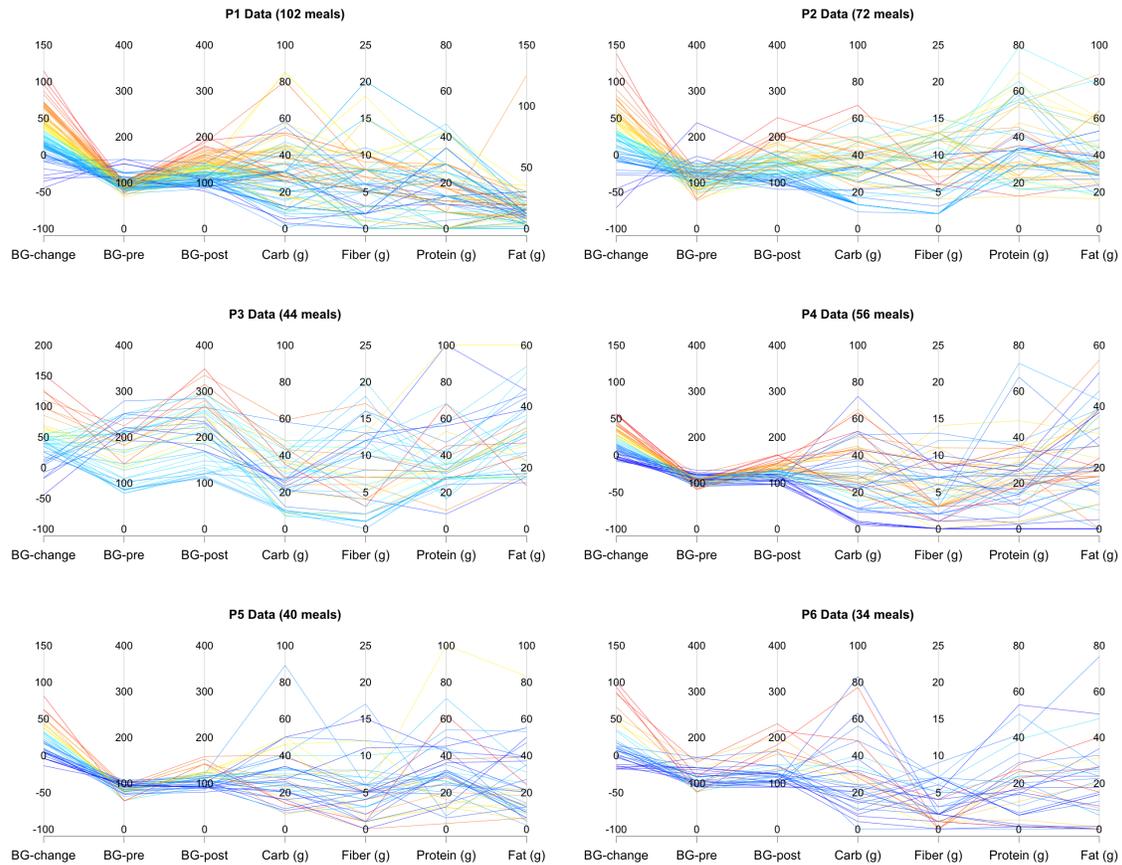

Parallel coordinate plots of behavioral-clinical self-monitoring data for all participants. Each line represents a single meal, and each vertical axis represents a behavioral or clinical variable (labeled at the bottom of each plot). The line color indicates the degree of glucose change, where red is highest and blue is lowest. These plots demonstrate the heterogeneity of behavioral-clinical patterns within and across participants.

**Supplementary Figure 2.**

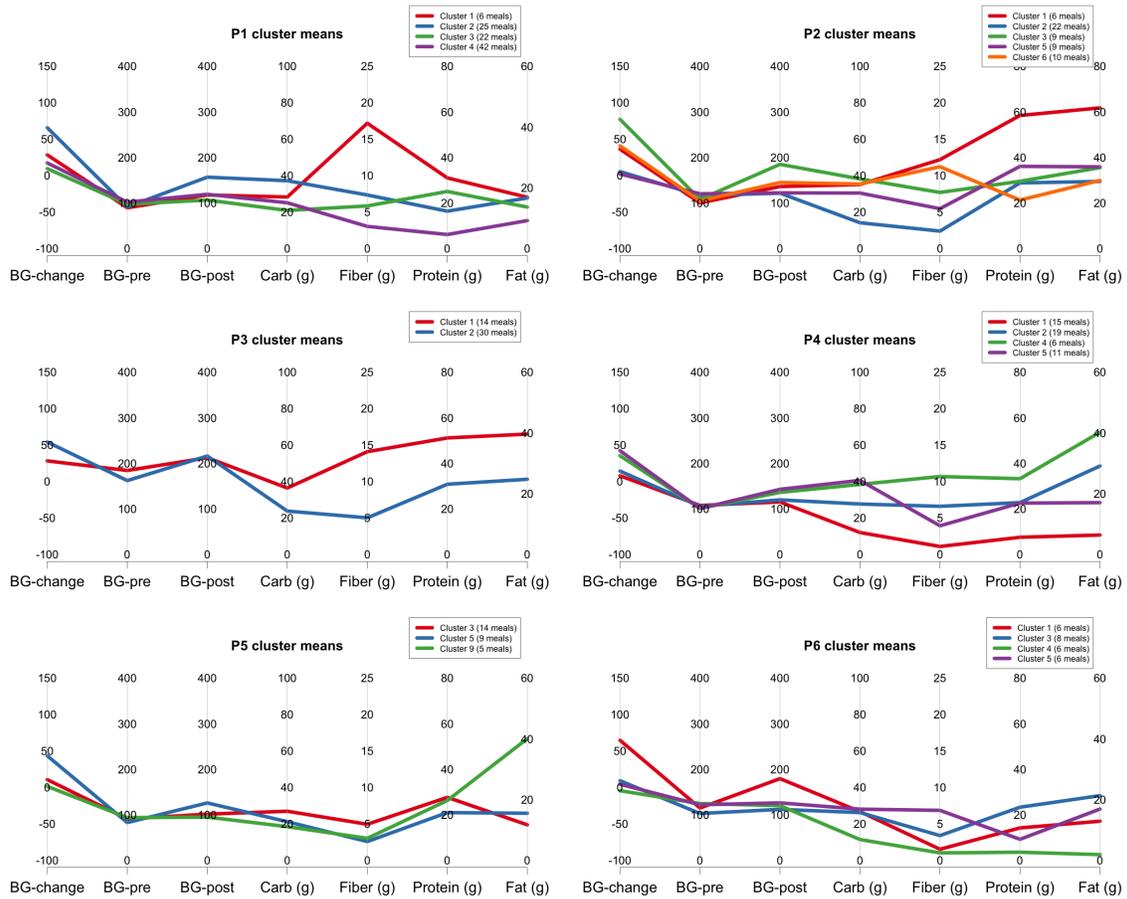

Parallel coordinate plots of cluster means from all participants. Each line represents the mean values for a single cluster. Clustering was performed using the blood glucose with grams of macronutrients feature set, and clusters with fewer than 5 meals were excluded. We observe that each participant exhibits different personal behavioral-clinical phenotypes.